\documentclass[conference,romanappendices,10pt]{IEEEtran}
\IEEEoverridecommandlockouts

\usepackage{algorithmic,amsmath,amssymb,amsthm,array,bbm,bibentry,cite,color,comment}
\usepackage{enumerate,enumitem,eurosym,float,graphicx,lettrine,mathrsfs,multirow,nomencl,import}
\usepackage{pict2e,psfrag,ragged2e,relsize,setspace,slashed,subfigure,supertabular,tabularx,url}
\usepackage[english]{babel}
\usepackage[T1]{fontenc}
\usepackage[utf8]{inputenc}

{		\newtheorem{theorem}{Theorem}}
{ 		}
{ 		}
{ 		}
{ 		}
{		}
{		}
{ 		\newtheorem{corollary}{Corollary}}

\newcommand{\setF}{\mathcal{F}}

\newcommand{\setL}{\mathcal{L}}

\newcommand{\diff}{\mathrm{d}}

\renewcommand{\Pr}{\mathbb{P}}

\renewcommand{\sc}{\textnormal{\tiny{SC}}}
\newcommand{\ut}{\textnormal{\tiny{UT}}}
\newcommand{\bh}{\textnormal{\tiny{BH}}}
\newcommand{\rmS}{\mathrm{S}}
\newcommand{\rmD}{\mathrm{D}}

\newcommand{\sinr}{\mathsf{SINR}}
\newcommand{\ase}{\mathsf{ASE}}
\newcommand{\aec}{\mathsf{AEC}}
\newcommand{\ee}{\mathsf{EE}}
\newcommand{\Psuc}{\mathsf{P}_{\mathrm{suc}}}
\newcommand{\Phit}{\mathsf{P}_{\mathrm{hit}}}

\title{Flexible Cache-Aided Networks with Backhauling}
\author{
	\IEEEauthorblockN{Italo Atzeni,$^{1}$ Marco Maso,$^{1}$ Imène Ghamnia,$^{2}$ Ejder Ba\c{s}tu\u{g},$^{2,3}$ and Mérouane Debbah$^{1,2}$}
	\IEEEauthorblockA{${}^{1}$Mathematical and Algorithmic Sciences Lab, France Research Center, Huawei Technologies France SASU \\
		${}^{2}$Large Networks and Systems Group (LANEAS), CentraleSupélec \\
		${}^{3}$Research Laboratory of Electronics, Massachusetts Institute of Technology (MIT) \\
		Email: \{italo.atzeni, marco.maso, merouane.debbah\}@huawei.com, imene.ghamnia@centralesupelec.fr, ejder@mit.edu \vspace{-1.0cm}}
	\thanks{This research has been supported in part by the ERC Starting Grant 305123 MORE (Advanced Mathematical Tools for Complex Network Engineering), the U.S. National Science Foundation under Grant CCF-1409228.}
}

\begin{document}
\IEEEoverridecommandlockouts
\maketitle
	
\begin{abstract}
Caching at the edge is a promising technique to cope with the increasing data demand in wireless networks. This paper analyzes the performance of cellular networks consisting of a tier macro-cell wireless backhaul nodes overlaid with a tier of cache-aided small cells. We consider both \emph{static} and \emph{dynamic} association policies for content delivery to the user terminals and analyze their performance. In particular, we derive closed-form expressions for the area spectral efficiency and the energy efficiency, which are used to optimize relevant design parameters such as the density of cache-aided small cells and the storage size. By means of this approach, we are able to draw useful design insights for the deployment of highly performing cache-aided tiered networks.
\vspace{-0.25cm}
\end{abstract}

%=========================================================================
\section{Introduction} \label{sec:intro}
%=========================================================================

A tremendous increase in smartphone usage during the last decade resulted in a $4000$-fold mobile data traffic explosion in cellular networks \cite{Cisco2016}. In this setting, mobile operators constantly look for innovative solutions to satisfy performance indicators such as higher spectral and energy efficiency, improved coverage, and lower end-to-end delays: this calls for the development of innovative techniques at both device and network level in the coming years \cite{Andrews2014Will}. Caching at the edge is regarded as one of the most promising approaches to cope with the increasing data demand \cite{Bas15}. This is particularly true for heterogeneous and tiered network layouts, for which the extent of the performance gains brought by caching has been extensively assessed in terms of several performance metrics of practical relevance. For instance, \cite{li2016optimization} and \cite{Liu16} focus on optimal probabilistic caching policies for enhanced content delivery, \cite{abd2017cache} studies the impact of caching on the coverage and on the delay experienced by the user terminals (UTs), and \cite{khreishah2016joint} jointly considers caching, routing, and channel assignment strategies for collaborative small cells~(SCs).

Along similar lines, this work analyzes the system performance of cache-aided tiered networks comprised of a tier of macro-cell backhaul (BH) nodes overlaid with a tier of non-cooperative SCs \cite{Mas17}. Each SC in the considered network has (limited) storage capabilities and pre-fetches popular files before the downlink transmission to its associated UTs. In this context, and differently from what has been previously proposed in the literature, our goal is to optimize relevant performance metrics for network planners and operators, and draw design guidelines for the deployment of highly performing cache-aided tiered networks. Using consolidated random spatial models to characterize the location of SCs, BHs and UTs \cite{Atz17}, we investigate the area spectral efficiency (ASE) gains brought by caching popular files in the considered network using tools from stochastic geometry. To increase the generality and relevance of our study, we examine both \emph{static} and \emph{dynamic} association policies for content delivery to the UTs. In particular, in the static case, UTs can only be associated with SCs, regardless of the occurrence of cache hit/miss events at the latter; conversely, in the dynamic case, UTs are associated with SCs in case of cache hit and with wireless BHs otherwise. 

The first contribution of this paper is the derivation of closed-form expressions and bounds for the ASE and the energy efficiency (EE) with the aforementioned UT association policies. Such expressions are subsequently used to optimize crucial design parameters such as SC density and storage size. Our findings highlight that the impact of caching on the ASE is minor as compared to that of the SC density, i.e., densifying the network is always more convenient than expanding the storage size at the SCs. Conversely, we show that increasing the storage size rather than the SC density is more effective in terms of EE of the network. In practice, the choice of investing on additional SCs or on more storage strongly depends on the goal of the network operator. Finally, it is worth observing that the performance of the considered network heavily hinges on the adopted UT association policy: a dynamic policy is always preferable whenever the network can support the larger complexity necessary for its effective implementation.

%=========================================================================
\section{System Model} \label{sec:syst}
%=========================================================================
\subsection{Network Model} \label{sec:syst_network}
%=========================================================================

Let us consider a set of mobile UTs in a two-tier ultra-dense network, where the lower tier comprises a set of non-cooperative SCs ensuring network coverage and the upper tier consists in a set of BHs connected to the internet. In our model, each SC is associated with only one UT and one BH. We assume that the spatial distribution of the network nodes follows the stationary, independently marked Poisson point processes (PPP) $\Phi \triangleq \big\{ \big( x, u(x), b(x) \big) \big\} \subset \mathbb{R}^{2} \times \mathbb{R}^{2} \times \mathbb{R}^{2}$. We use $\Phi_{\sc} \triangleq \{ x \}$ to denote the PPP of the SCs with spatial density $\lambda$ (measured in SCs/m$^{2}$), with isotropic marks $\Phi_{\ut} \triangleq u(\Phi_{\sc}) = \{ u(x) \}_{x \in \Phi_{\sc}}$ and $\Phi_{\bh} \triangleq b(\Phi_{\sc}) = \{ b(x) \}_{x \in \Phi_{\sc}}$ representing the UTs and the BHs, respectively; evidently, $\Phi_{\ut}$ and $\Phi_{\bh}$ are dependent on $\Phi_{\sc}$ and have the same spatial density $\lambda$. In this setting, the employment of PPPs allows to capture the randomness of practical ultra-dense network deployments and, at the same time, obtain precise and tractable expressions for the system-level performance metrics \cite{Atz17,And11,Atz17b}. Let $r_{y z} \triangleq \| y - z \|$ denote the distance between nodes $y,z \in \Phi$: the distances of the UTs and the BHs from their associated SCs are assumed fixed and are denoted by $R_{\ut} \triangleq r_{x u(x)}$ and $R_{\bh} \triangleq r_{b(x) x}$, $\forall x \in \Phi_{\sc}$, respectively, with $R_{\bh} > R_{\ut}$.

%=========================================================================
\subsection{Caching Model and UT Association} \label{sec:syst_caching}
%=========================================================================

%\footnote{Note that files with different lengths can be divided into chunks of equal length.}
Let $\setF \triangleq \{ f_{i} \}_{i=1}^{F}$ be a $F$-sized subset of all the files available in the internet, entirely accessible by each BH. Without loss of generality, and for simplicity in the notation, we assume that all $F$ files have identical lengths. In the following, we refer to $\setF$ as \textit{file catalog}. Each UT requests files from $\setF$ with probability denoted by $\mathcal{P} \triangleq \lbrace p_{1}, p_{2}, \ldots, p_{F} \rbrace$, with $\sum_{i=1}^{F} p_{i} = 1$, which hinges on the files popularity over the whole network. 

To offload the overlaying BH infrastructure, each non-cooperative SC $x$ is equipped with a \textit{storage unit} ${\Delta_{x}}$ of size $S \leq F$ (measured in files/SC). Suppose that, at a given time instant, UT $u(x)$ is interested in downloading file $f_{i} \in \setF$: if $f_{i}$ is cached at its associated SC $x$, i.e., $f_{i} \in {\Delta_{x}}$, we have a \textit{cache hit} event; conversely, if  $f_{i} \notin {\Delta_{x}}$, we have a \textit{cache miss} event.  In this regard, we use $\slashed{\Delta}_{x}$ to denote a cache miss event at $x$. Accordingly, we introduce the indicator function
\begin{align}
\mathbbm{1}_{\slashed{\Delta}_{x}} \triangleq \begin{cases}
1, & \mathrm{if} \ \slashed{\Delta}_{x} \\
0, & \mathrm{otherwise}
\end{cases}
\end{align}
where its logical complement is denoted by $\mathbbm{1}_{\Delta_{x}}$. 

The effectiveness of the adopted cache-aided approach depends on the probability that any file requested by a given UT is cached at its serving SC, which is referred to as \textit{cache hit probability} and is denoted by $\Phit$. This quantity typically depends on the adopted caching policy, content placement, and distribution of the UT requests. Further details on these aspects are unnecessary for the scope of our work and we refer to \cite{Bas15} and references therein for further details.   

Now, let $R=TB \in \mathbb{R}$ be the time-frequency resource used by the network for delivering one file from the BH tier to the UTs, with $T \in \mathbb{R}$ (resp. $B \in \mathbb{R}$) defined as the amount of necessary time (resp. frequency) resources to perform this operation. Two UT association policies are considered in this work, i.e., \emph{static} and \emph{dynamic}. When the static UT association is adopted, operations in the two tiers occur on orthogonal resources. In this case, the UTs can only be served by their associated SCs, and the BH-to-SC/SC-to-UT links occupy two $\frac{R}{2}$-sized resources: for instance, each link may use the entirety of $B$ for half of the time (i.e., $\frac{T}{2}$) or the entirety of $T$ for half of the bandwidth (i.e., $\frac{B}{2}$). 
\begin{figure}[!t]
	\centering
	{\def\svgwidth{.85\columnwidth} \import{.}{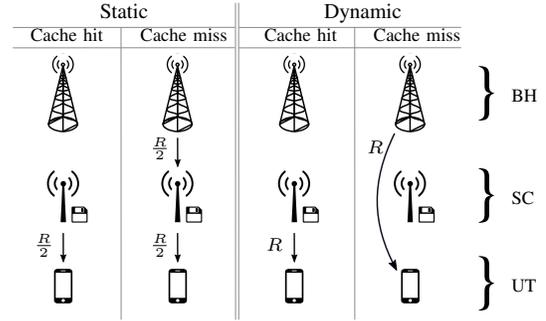}}
	\vspace{-2mm}
	\caption{System model for static and dynamic UT association.} \label{fig:system_model} \vspace{-3mm}
\end{figure}
Conversely, under dynamic UT association, operations in the two tiers occur on the same resource. In this case, the UTs are served by either the SCs or the BHs depending on the availability of the requested file at the SCs, and the SC-to-UT/BH-to-UT links occupy a $R$-sized resource. In practice, the second approach leads to a more efficient use of the time-frequency resource at the cost of more sophisticated inter-tier coordination and possibly signaling. For notational simplicity, in the rest of the paper we differentiate between static and dynamic UT association by means of the super-indices $\rmS$ and $\rmD$, respectively. 

The adopted UT association policy has a strong impact on the network response to a cache hit/miss event when UT $u(x)$ requests file $f_{i}$. In particular:
\begin{itemize}
	\item[\textit{1)}] \textit{Static}: If a cache hit occurs, then SC $x$ transmits $f_{i}$ to $u(x)$. Conversely, if a cache miss occurs, first $x$ retrieves $f_{i}$ from its associated BH $b(x)$, then transmits it to UT $u(x)$. In both cases, an $\frac{R}{2}$-sized resource is used by the SC-to-UT link due to the static UT association;
	\item[\textit{2)}] \textit{Dynamic}: If a cache hit occurs, then $x$ transmits $f_{i}$ to $u(x)$. Conversely, if a cache miss occurs, $u(x)$ can bypass $x$ and connect directly to $b(x)$ to download $f_{i}$. In both cases, the whole resource $R$ is used by either the SC-to-UT or the BH-to-UT link.
\end{itemize}
A graphical representation of these operations, and their impact on the network resource usage, is provided in Figure \ref{fig:system_model}.

\vspace{-1mm}

%=========================================================================
\subsection{Channel Model} \label{sec:syst_ch}
%=========================================================================

%\footnote{Our analysis can be extended to the case of multi-antenna nodes using the framework in \cite{Atz17}.}
In this work, all nodes have a single transmit/receive antenna. In addition, we assume that the SCs and the BHs transmit with powers $\rho_{\sc}$ and $\rho_{\bh}$ (measured in W), respectively. The propagation through the wireless channel combines pathloss attenuation and small-scale fading. For the former, we adopt the standard power-law pathloss model and define the pathloss function $\ell(y,z) \triangleq r_{y z}^{- \alpha}$ between nodes $y,z \in \Phi$, with pathloss exponent $\alpha > 2$. For the latter, we use $h_{y z}$ to denote the channel power fading gain between nodes $y,z \in \Phi$:  all channels are assumed to be subject to Rayleigh fading and, hence, $h_{y z} \sim \exp(1)$.\footnote{For more involved types of small-scale fading, e.g., Nakagami$-m$ fading, we refer to \cite{Atz17b}.}
The signal-to-interference-plus-noise ratio (SINR) at SC $x$ reads as
\begin{align}
\sinr_{x} \triangleq \frac{\rho_{\bh} R_{\bh}^{-\alpha} h_{b(x) x}}{I_{x} + \sigma^{2}}
\end{align}
where $I_{x}$ is the overall interference power at SC $x$ and $\sigma^{2}$ is the additive noise power. Likewise, the SINR at UT $u(x)$ reads~as
\begin{align}
\sinr_{u(x)} \triangleq \frac{\rho_{\sc} R_{\ut}^{-\alpha} h_{x u(x)}}{I_{u(x)} + \sigma^{2}}
\end{align}
where $I_{u(x)}$ is the overall interference power at $u(x)$. For static UT association, we have $I_{x} = I_{x}^{(\rmS)}$ and $I_{u(x)} = I_{u(x)}^{(\rmS)}$, with
\begin{align}
\label{eq:I_x_S} I_{x}^{(\rmS)} & \triangleq \sum_{y \in \Phi_{\sc} \backslash \{ x \}} \rho_{\bh} r_{b(y) x}^{-\alpha} h_{b(y) x} \mathbbm{1}_{\slashed{\Delta}_{y}} \\
\label{eq:I_ux_S} I_{u(x)}^{(\rmS)} & \triangleq \sum_{y \in \Phi_{\sc} \backslash \{ x \}} \rho_{\sc} r_{y u(x)}^{-\alpha} h_{y u(x)}
\end{align}
respectively. On the other hand, in case of dynamic UT association, we are only interested in the interference at the UT and we thus have $I_{u(x)} = I_{u(x)}^{(\rmD)}$, with
\begin{align}
\nonumber I_{u(x)}^{(\rmD)} \triangleq \sum_{y \in \Phi_{\sc} \backslash \{ x \}} & \big( \rho_{\sc} r_{y u(x)}^{-\alpha} h_{y u(x)} \mathbbm{1}_{\Delta_{y}} \\
\label{eq:I_ux_D} & + \rho_{\bh} r_{b(y) u(x)}^{-\alpha} h_{b(y) u(x)} \mathbbm{1}_{\slashed{\Delta}_{y}} \big).
\end{align}

%=========================================================================
\section{Performance Analysis} \label{sec:perf}
%=========================================================================

The system performance is closely related to the probability of the UTs successfully receiving the requested content, which is termed as \textit{success probability}. This depends on two factors, i.e., the cache hit probability (cf. Section~\ref{sec:syst_caching}) and the SINRs of the file transmissions (cf. Section~\ref{sec:syst_ch}). Thus, in the following, we derive the success probability using both the considered UT association policies. %To this end, we focus on the file transmissions in the downlink and neglect the overhead induced by the file requests by the UTs in the uplink.

Our analysis focuses on a randomly chosen SC $x$, and its marks $u(x)$ and $b(x)$, referred to as \textit{typical SC}, \textit{typical UT} and \textit{typical BH}, respectively: due to Slivnyak's theorem and to the stationarity of $\Phi$, these nodes are representative of the whole network \cite{Atz17}. For a given SINR threshold $\theta$, we assume that a file is successfully received by the typical UT if $\sinr_{x} > \theta \land \sinr_{u(x)} > \theta$ and $\sinr_{u(x)} > \theta$ for static and dynamic UT association, respectively. The success probability is formalized in the following theorem, where we use the notation
\begin{align}
\Upsilon (z) & \triangleq \frac{\pi z^{\frac{2}{\alpha}} \csc \big( \tfrac{2 \pi}{\alpha} \big)}{\alpha}.
\end{align}
\begin{theorem} \label{th:Psuc} \rm{
For static UT association, the success probability is given by
\begin{align}
\nonumber & \Psuc^{(\rmS)} (\theta) \triangleq \exp \big( - \theta \tfrac{\sigma^{2}}{\rho_{\sc}} R_{\ut}^{\alpha} \big) \setL_{I_{u(x)}}^{(\rmS)} \big( \theta \rho_{\sc}^{-1} R_{\ut}^{\alpha} \big) \\
\label{eq:Psuc_S} & \times \big( \Phit + (1 - \Phit) \exp \big( - \theta \tfrac{\sigma^{2}}{\rho_{\bh}} R_{\bh}^{\alpha} \big) \setL_{I_{x}}^{(\rmS)} \big( \theta \rho_{\bh}^{-1} R_{\bh}^{\alpha} \big) \big)
\end{align}
where
\begin{align}
\label{eq:L_I_x_S} \setL_{I_{x}}^{(\rmS)}(s) & \triangleq \exp \big( - 2 \pi \lambda (1 - \Phit) \Upsilon (\rho_{\bh} s) \big) \\
\label{eq:L_I_ux_S} \setL_{I_{u(x)}}^{(\rmS)}(s) & \triangleq \exp \big( - 2 \pi \lambda \Upsilon (\rho_{\sc} s) \big)
\end{align}
are the Laplace transforms of the interference terms $I_{x}^{(\rmS)}$ and $I_{u(x)}^{(\rmS)}$ in \eqref{eq:I_x_S}--\eqref{eq:I_ux_S}, respectively. For dynamic UT association, the success probability is given by
\begin{align}
\nonumber & \hspace{-1mm} \Psuc^{(\rmD)} (\theta) \! \triangleq \! \Phit \exp \big( \! - \theta \tfrac{\sigma^{2}}{\rho_{\sc}} R_{\ut}^{\alpha} \big) \setL_{I_{u(x)}}^{(\rmD)} \big( \theta \rho_{\sc}^{-1} R_{\ut}^{\alpha} \big) + (1 - \Phit) \\
\label{eq:Psuc_D} & \hspace{-2mm} \times \exp \big( \! - \theta \tfrac{\sigma^{2}}{\rho_{\bh}} R_{\bh}^{\alpha} \big) \! \int_{0}^{2 \pi} \! \! \setL_{I_{u(x)}}^{(\rmD)} \big( \theta \rho_{\bh}^{-1} \Omega(R_{\ut},R_{\bh},\phi) \big) \frac{\diff \phi}{2 \pi}
\end{align}
with $\Omega(R_{\ut},R_{\bh},\phi) \! \triangleq \! \big( R_{\ut}^{2} \! + \! R_{\bh}^{2} \! + \! 2 R_{\ut} R_{\bh} \cos \phi \big)^{\frac{\alpha}{2}}$ and where

\newpage

$ $ \vspace{-9mm}

\begin{align}
\nonumber \setL_{I_{u(x)}}^{(\rmD)}(s) \triangleq \exp & \big( - 2 \pi \lambda \Phit \Upsilon(\rho_{\sc} s) \big) \\
\label{eq:L_I_ux_D} & \times \exp \big( - 2 \pi \lambda (1 - \Phit) \Upsilon(\rho_{\bh} s) \big)
\end{align}
is the Laplace transform of the interference term $I_{u(x)}^{(\rmD)}$ in \eqref{eq:I_ux_D}.
}
\end{theorem}
\begin{IEEEproof}
See Appendix~\ref{sec:app_Psuc}.
\end{IEEEproof} \vspace{2mm}
\noindent Since a closed-form expression for the success probability $\Psuc^{(\rmD)} (\theta)$ in \eqref{eq:Psuc_D} is not available, we provide a useful lower bound in the following corollary.

\begin{corollary} \label{cor:Psuc} \rm{
For dynamic UT association, the success probability can be lower-bounded as
\begin{align}
\nonumber & \Psuc^{(\rmD)} (\theta) \geq \Phit \exp \big( - \theta \tfrac{\sigma^{2}}{\rho_{\sc}} R_{\ut}^{\alpha} \big) \setL_{I_{u(x)}}^{(\rmD)} \big( \theta \rho_{\sc}^{-1} R_{\ut}^{\alpha} \big) \\
\label{eq:Psuc_D_lb} & + (1 - \Phit) \exp \big( - \theta \tfrac{\sigma^{2}}{\rho_{\bh}} R_{\bh}^{\alpha} \big) \setL_{I_{x}}^{(\rmD)} \big( \theta \rho_{\bh}^{-1} \big( R_{\bh} + R_{\ut} \big)^{\alpha} \big).
\end{align}
}
\end{corollary}

\begin{IEEEproof}
The lower bound is obtained by considering the worst-case distance between BH and UT, i.e., $R_{\bh} + R_{\ut}$ (cf. \cite{Atz17}).
\end{IEEEproof} \vspace{2mm}

\begin{figure}[t!]
	\centering
	\includegraphics[scale=0.82]{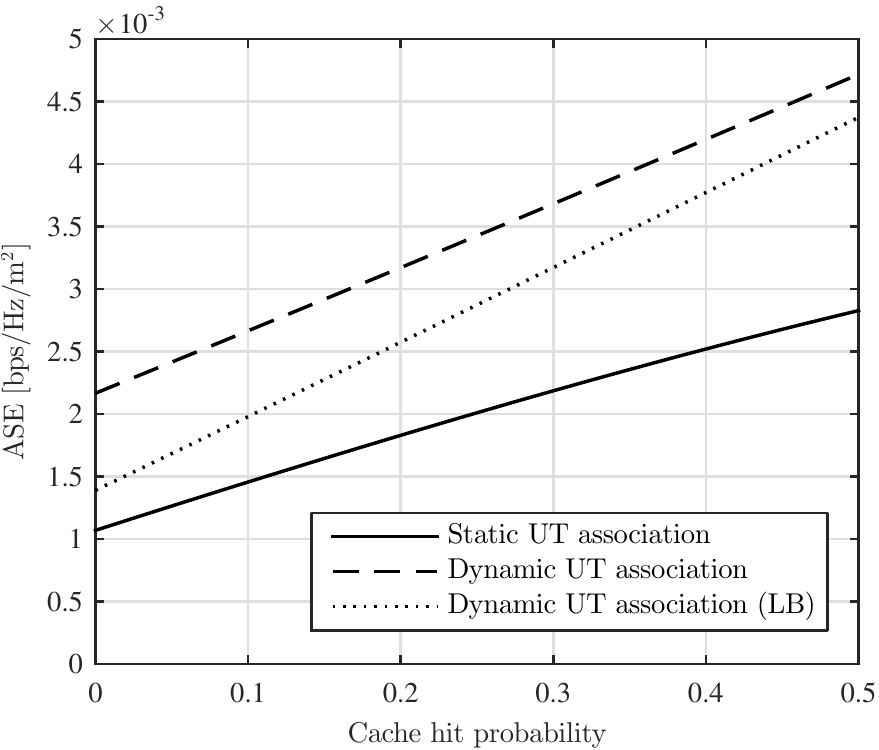}
	\vspace{-2mm}
	\caption{Achievable ASE as a function of the cache hit probability.} \label{fig:Psuc} \vspace{-3mm}
\end{figure}

The expressions above can be used to obtain other useful performance metrics. In this paper, we focus on the achievable ASE (measured in bps/Hz/m$^{2}$), which is readily obtained as
\begin{align}
\label{eq:ASE_S} \ase^{(\rmS)}(\theta, \lambda) & \triangleq \frac{1}{2} \lambda \Psuc^{(\rmS)} (\theta) \log_{2}(1 + \theta) \\
\label{eq:ASE_D} \ase^{(\rmD)}(\theta, \lambda) & \triangleq \lambda \Psuc^{(\rmD)} (\theta) \log_{2}(1 + \theta)
\end{align}
for static and dynamic UT association, respectively. At this stage, it is worth observing that, in practice, the distances $R_{\ut}$ and $R_{\bh}$ depend on the SC density $\lambda$. Thus, in the following we set $R_{\ut} = \frac{\beta_{\ut}}{2 \sqrt{\lambda}}$ and $R_{\bh} = \frac{\beta_{\bh}}{2 \sqrt{\lambda}}$, with $\beta_{\bh} > \beta_{\ut}$; observe that $\frac{1}{2 \sqrt{\lambda}}$ represents the average distance between nodes in a PPP with spatial density $\lambda$. Figure~\ref{fig:Psuc} illustrates the achievable ASEs as defined in \eqref{eq:ASE_S}--\eqref{eq:ASE_D} against the cache hit probability $\Phit$, with $\lambda = 10^{-2}$~SCs/m$^{2}$, $\alpha=4$, $\theta = 1$, $\sigma^{2} = 0$ (i.e., interference limited case), $\rho_{\sc} = 0.5$~W, $\rho_{\bh} = 1$~W, $\beta_{\sc} = 0.5$, and $\beta_{\bh} = 1$ (the same parameters will be used in Section~\ref{sec:opt_1}). The beneficial impact of caching is evident: as compared with the cache-free setup (which corresponds to $\Phit=0$), the achievable ASE with $\Phit=0.25$ (resp. $\Phit=0.5$) is increased by $188\%$ (resp. $264\%$) for dynamic UT association and by $158\%$ (resp. $218\%$) for static UT association. Furthermore, we observe that the dynamic UT association always outperforms the static UT association. Lastly, the lower bound on \eqref{eq:ASE_D}, derived using Corollary~\ref{cor:Psuc}, is increasingly tight as $\Phit$ grows large.

%=========================================================================
\section{Cache-Aided SC Network Deployment} \label{sec:opt}
%========================================================================

In this section, we look at possible design problems of interest to network operators and we aim at providing simple design guidelines for the deployment of cache-aided SC networks. 
As design parameters, we focus on the SC density $\lambda \in [\lambda^{(\min)}, \lambda^{(\max)}]$ and the storage size $S \in [0, S^{(\max)}]$, with $S^{(\max)} \leq F$: therefore, we make the dependence on these parameters explicit in our performance metrics. For instance, the minimum SC density $\lambda^{(\min)}$ can be chosen to ensure an interference-limited system performance (i.e., the background noise becomes negligible), whereas the maximum SC density $\lambda^{(\max)}$ can be due to spatial limitations; similarly, the maximum storage size $S^{(\max)}$ may arise from practical constraints regarding space or power consumption at the SCs.

In our analysis, we consider the simplest case of uniform files popularity, i.e., $p_{i} = \frac{1}{F}$, $i=1, \ldots, F$, with $\Phit(S) = \frac{S}{F}$, where we have made the dependence of the cache hit probability on $S$ explicit. Note that, for the more realistic case of decreasing files popularity, i.e., $p_{1} > p_{2} > \ldots > p_{F}$, expanding the storage will have an increasingly less significant impact on the system performance.

\vspace{-1mm}

%=========================================================================
\subsection{Deployment Cost} \label{sec:opt_1}
%=========================================================================

Suppose that a network operator has a fixed monetary budget per m$^{2}$ $c$ (measured in \$/m$^{2}$) to deploy a cache-aided SC network. Let $p_{\lambda}$ denote the SC price (measured in \$/SC) and let $p_{S}$ denote the storage price (measured in \$/file). The following simple question naturally arises in this context: \textit{is it better to invest in deploying SCs or storage?} The answer hinges on the choice of the performance metric and can be found by means of an appropriate optimization. For instance, to maximize the achievable ASE in \eqref{eq:ASE_S}--\eqref{eq:ASE_D}, the answer is given by the solution of the following optimization problem:
\begin{align} \label{eq:opt1} \tag{$\mathrm{P}1$}
\begin{array}{ccll} %\vspace{2mm}
\displaystyle \max_{\lambda, S} & & \ase(\lambda, S) \\
\displaystyle \mathrm{s.t.} & & p_{\lambda} \lambda + p_{S} \lambda S \leq c \\
							& & \lambda \in [\lambda^{(\min)}, \lambda^{(\max)}] \\
							& & S \in [0, S^{(\max)}].
\end{array}
\end{align}
Alternatively, the system performance can be measured in terms of its EE. Before formalizing the corresponding problem, some definitions are in order. Let
\begin{align}
E_{\mathrm{tot}}(S) \triangleq \Phit(S) E_{\mathrm{hit}} + \big( 1 - \Phit(S) \big) E_{\mathrm{miss}}
\end{align}
be the total energy consumed to provide a content to a UT (measured in J), where $E_{\mathrm{hit}}$ is the energy needed by the SC to retrieve a cached file from its storage unit and transmit it to the UT in case of cache hit, and $E_{\mathrm{miss}}$ is the energy needed by the BH to retrieve a non-cached file from the internet and transmit it to the UT in case of cache miss (via the SC for static UT association and directly for dynamic UT association). 
\begin{figure}[t!]
	\centering
	\includegraphics[scale=0.82]{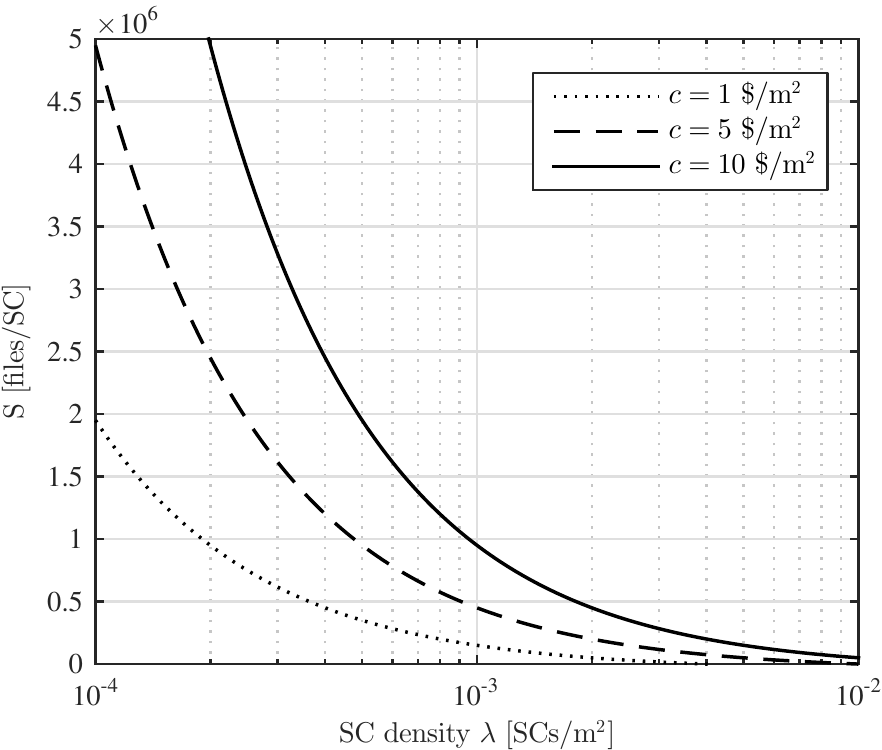}
	\vspace{-2mm}
	\caption{Feasible solution set of problems \eqref{eq:opt1}--\eqref{eq:opt2} for different values of the monetary budget~$c$.} \vspace{-3mm} \label{fig:Pareto} 
\end{figure}
Moreover, we define the area energy consumption (measured in J/m$^{2}$) as $\aec (\lambda,S) \triangleq \lambda E_{\mathrm{tot}}(S)$ and the EE (measured in bit/J) as
%
%
%\vspace{-3mm}
\begin{align}
\label{eq:EE} \ee (\lambda, S) \triangleq \frac{\ase(\lambda, S)}{\aec (\lambda,S)}.
\end{align}
Hence, the corresponding optimization problem is given by:
\begin{align} \label{eq:opt2} \tag{$\mathrm{P}2$}
\begin{array}{ccll} %\vspace{2mm}
\displaystyle \max_{\lambda, S} & & \ee(\lambda, S) \\
\displaystyle \mathrm{s.t.} & & p_{\lambda} \lambda + p_{S} \lambda S \leq c \\
							& & \lambda \in [\lambda^{(\min)}, \lambda^{(\max)}] \\
							& & S \in [0, S^{(\max)}].
\end{array}
\end{align}

Using the same parameters as in Section~\ref{sec:perf}, we additionally set $F=10^{7}$~files, $\lambda^{(\min)} = 10^{-4}$~SCs/m$^{2}$, $\lambda^{(\max)} = 10^{-2}$~SCs/m$^{2}$, $S^{(\max)} = 5 \times 10^{6}$~files/SC, $p_{\lambda}=250$~\$/SC, $p_{\lambda}=0.005$~\$/file, $E_{\mathrm{hit}} = 1$~J, and $E_{\mathrm{miss}}=10$~J. Assuming $p_{\lambda} \lambda^{\star} + p_{S} \lambda^{\star} S^{\star} = c$, which excludes the trivial solution $\{ \lambda^{\star}, S^{\star} \} = \{ \lambda^{(\max)}, S^{(\max)} \}$, Figure~\ref{fig:Pareto} illustrates the feasible solution set of the considered problems with $c=\{ 1, 2.5, 5\}$~\$/m$^{2}$.

\begin{figure}[t!]
\centering
\includegraphics[scale=0.82]{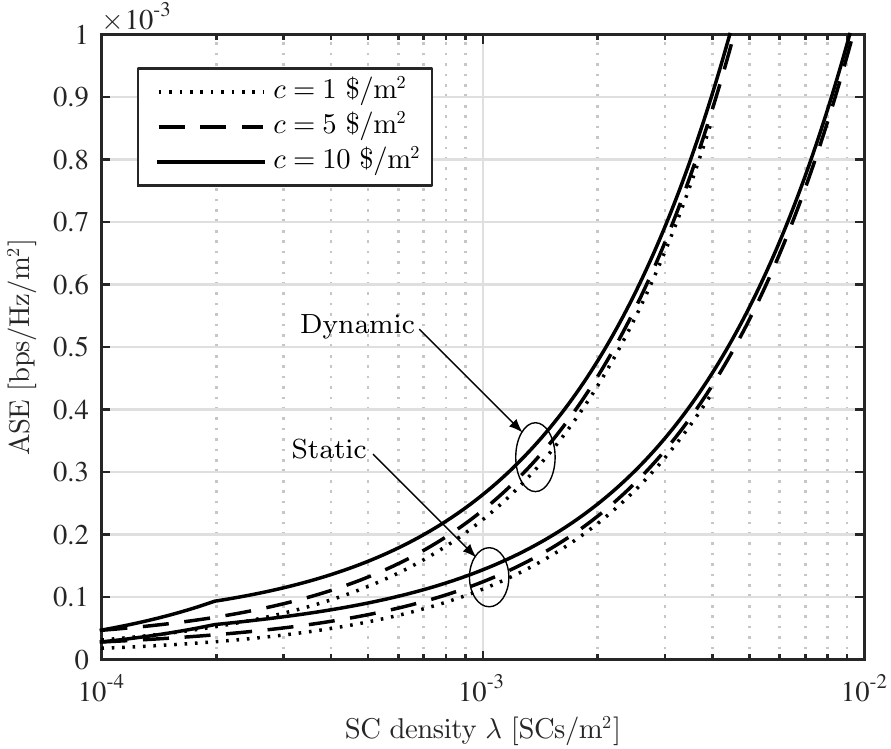}
\vspace{-2mm}
\caption{Achievable ASE as a function of the SC density.} \label{fig:ASE_opt} \vspace{-3mm}
\end{figure}

\begin{figure}[t!]
\centering
\includegraphics[scale=0.82]{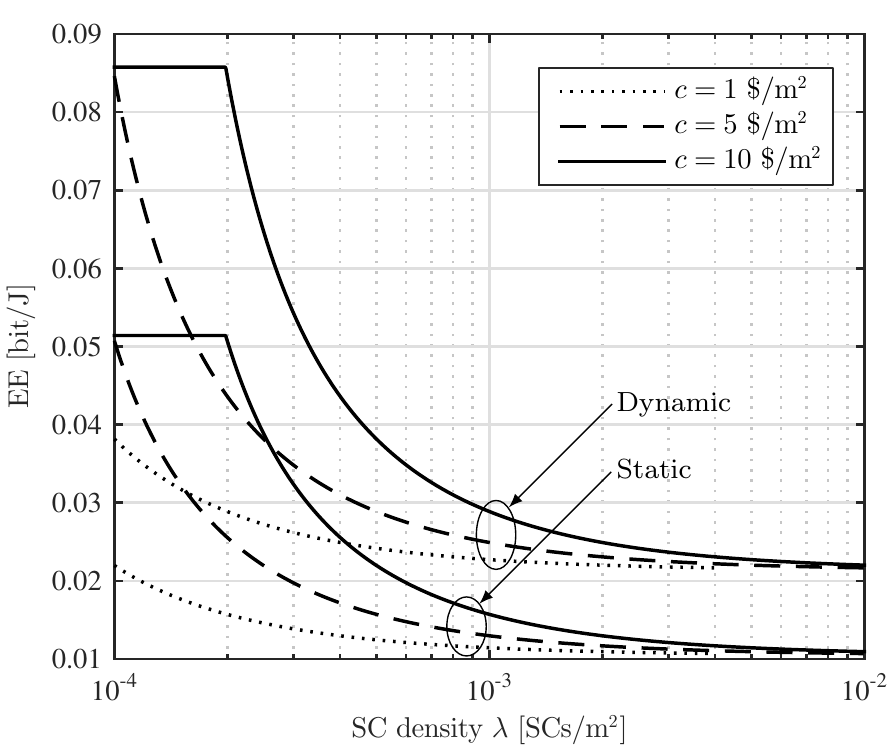}
\vspace{-2mm}
\caption{EE as a function of the SC density.} \label{fig:EE_opt} \vspace{-3mm}
\end{figure}

Figures~\ref{fig:ASE_opt} and \ref{fig:EE_opt} plot the achievable ASE and the EE, respectively, against the SC density $\lambda$, where each value of $\lambda$ yields a value of $S$ (and thus of $\Phit$) according to the constraint on the monetary budget (cf. Figure~\ref{fig:Pareto}). The achievable ASE in Figure~\ref{fig:ASE_opt} is monotonically increasing in $\lambda$: this means that, to maximize the network ASE, one should invest in deploying more SCs (even if that means $S=0$), whereas $S$ should be increased only if there is remaining budget. Therefore, the solution of problem \eqref{eq:opt1} is given by
\begin{equation}
\lambda^{\star} = \min \bigg\{ \frac{c}{p_{\lambda}}, \lambda^{(\max)} \bigg\}, \quad S^{\star} = \frac{1}{p_{S}} \bigg( \frac{c}{\lambda^{\star}} - p_{\lambda} \bigg).
\end{equation}
On the contrary, the EE in Figure~\ref{fig:EE_opt} is monotonically decreasing in $\lambda$: this means that, to maximize the network EE, one should invest in expanding the storage as much as possible (even if that means $\lambda = \lambda^{(\min)}$), whereas $\lambda$ should be increased only if there is remaining budget. Hence, the solution of problem \eqref{eq:opt2} reads as
\begin{equation}
S^{\star} = \min \bigg\{ \frac{1}{p_{S}} \bigg( \frac{c}{\lambda^{(\min)}} - p_{\lambda} \bigg), S^{(\max)} \bigg\}, \quad \lambda^{\star} = \frac{c}{p_{\lambda} + p_{S} S^{\star}}.
\end{equation}
In practice, increasing the SCs caching capabilities seems to be very useful for increasing the EE. Conversely, the impact of additional storage on the ASE is minor if the operator disposes of additional resources to increase the SC density. These results suggest that the answer to the question initially posed in this section, i.e., the choice of investing on additional SCs or on more storage, is not unique but strongly hinges on the ultimate goal of the network operator. Finally, we observe that a dynamic UT association policy is always preferable in terms of network performance whenever the larger complexity necessary to implement it can be afforded. 

\vspace{-1mm}

%=========================================================================
\section{Conclusions} \label{sec:concl}
%=========================================================================
In this work, we study the performance of a tiered network where a tier of macro-cell wireless backhaul nodes is overlaid with a tier of cache-aided SCs. To frame a more realistic scenario, we consider both \emph{static} and \emph{dynamic} UT association policies. Building on random spatial models and using tools from stochastic geometry, we derive closed-form expressions and bounds for the achievable ASE for both UT association policies. Then, we study suitable optimization problems to identify deployment strategies for maximizing the ASE and EE of the network. Our findings show that the ASE is maximized by first increasing the SC density and then the storage size at the SCs, whereas an opposite approach should be adopted for maximizing the EE.

\vspace{-1mm}

\appendices

%=========================================================================
\section{Proof of Theorem~\ref{th:Psuc}} \label{sec:app_Psuc}
%=========================================================================

Due to space limitations, we provide only a sketch of the proof. Considering static UT association, we have $\Psuc^{(\rmS)} (\theta) = \Phit \Pr[\sinr_{u(x)} > \theta] + (1 - \Phit) \Pr[\sinr_{u(x)} > \theta] \Pr[\sinr_{x} > \theta]$ and, building on \cite{And11,Mas17}, we obtain \vspace{-1mm}
\begin{align}
\Pr[\sinr_{u(x)} > \theta] & = \exp \big( - \theta \tfrac{\sigma^{2}}{\rho_{\sc}} R_{\ut}^{\alpha} \big) \setL_{I_{x}}^{(\rmS)} \big( \theta \rho_{\sc}^{-1} R_{\ut}^{\alpha} \big) \\
\Pr[\sinr_{x} > \theta] & = \exp \big( - \theta \tfrac{\sigma^{2}}{\rho_{\bh}} R_{\bh}^{\alpha} \big) \setL_{I_{x}}^{(\rmS)} \big( \theta \rho_{\bh}^{-1} R_{\bh}^{\alpha} \big)
\end{align}
with $\setL_{I_{u(x)}}^{(\rmS)}(s)$ and $\setL_{I_{x}}^{(\rmS)}(s)$ defined in \eqref{eq:L_I_x_S}--\eqref{eq:L_I_ux_S}. Likewise, for dynamic UT association, we have $\Psuc^{(\rmD)} (\theta) = \Pr[\sinr_{u(x)} > \theta]$ where (cf. \cite{And11,Mas17}) \vspace{-1mm}
\begin{align}
\nonumber \Pr[\sinr_{u(x)} > \theta] & = \exp \big( - \theta \tfrac{\sigma^{2}}{\rho_{\sc}} R_{\ut}^{\alpha} \big) \setL_{I_{u(x)}}^{(\rmS)} \big( \theta \rho_{\sc}^{-1} R_{\ut}^{\alpha} \big) \mathbbm{1}_{\Delta_{x}} \\
+ \exp & \big( - \theta \tfrac{\sigma^{2}}{\rho_{\bh}} R_{\bh}^{\alpha} \big) \setL_{I_{u(x)}}^{(\rmS)} \big( \theta \rho_{\bh}^{-1} R_{\bh}^{\alpha} \big) \mathbbm{1}_{\slashed{\Delta}_{x}}
\end{align}
with $\setL_{I_{u(x)}}^{(\rmD)}(s)$ defined in \eqref{eq:L_I_ux_D}. \hfill \IEEEQED

\addcontentsline{toc}{chapter}{References}
\bibliographystyle{IEEEtran}
\bibliography{IEEEabrv,ref_Huawei}

\end{document}